\documentclass[sn-mathphys-num]{sn-jnl}

\usepackage{graphicx}%
\usepackage{multirow}%
\usepackage{amsmath,amssymb,amsfonts}%
\usepackage{amsthm}%
\usepackage{mathrsfs}%
\usepackage[title]{appendix}%
\usepackage{xcolor}%
\usepackage{textcomp}%
\usepackage{manyfoot}%
\usepackage{booktabs}%
\usepackage{algorithm}%
\usepackage{algorithmicx}%
\usepackage{algpseudocode}%
\usepackage{listings}%

\theoremstyle{thmstyleone}%
\theoremstyle{thmstyletwo}%

\theoremstyle{thmstylethree}%

\begin{document}

\title[Article Title]{Radio Detection of ultra-high-energy Cosmic-Ray Air Showers}


\author*[1,2]{\fnm{Frank G.} \sur{Schr\"oder}}\email{frank.schroeder@kit.edu}

\affil[1]{\orgdiv{Bartol Research Institute, Department of Physics and Astronomy}, \orgname{University of Delaware}, \orgaddress{\street{Sharp Lab, 104 The Green}, \city{Newark}, \postcode{19716}, \state{Delaware}, \country{United States of America}}}

\affil[2]{\orgdiv{Institute for Astroparticle Physics}, \orgname{Karlsruhe Institute of Technology (KIT)}, \orgaddress{\street{Hermann-von-Helmholtz-Platz 1}, \postcode{76344} \city{Eggenstein-Leopoldshafen}, \country{Germany}}}


\abstract{Radio antennas have become a standard tool for the detection of cosmic-ray air showers in the energy range above $10^{16}\,$eV. 
The radio signal of these air showers is generated mostly due to the deflection of electrons and positrons in the geomagnetic field, and contains information about the energy and the depth of the maximum of the air showers.
Unlike the traditional air-Cherenkov and air-fluorescence techniques for the electromagnetic shower component, radio detection is not restricted to clear nights, and recent experiments have demonstrated that the measurement accuracy can compete with these traditional techniques.
Numerous particle detector arrays for air showers have thus been or will be complemented by radio antennas. 
In particular when combined with muon detectors, the complementary information provided by the radio antennas can enhance the total accuracy for the arrival direction, energy and mass of the primary cosmic rays. 
Digitization and computational techniques have been crucial for this recent progress, and radio detection will play an important role in next-generation experiments for ultra-high-energy cosmic rays.
Moreover, stand-alone radio experiments are under development and will search for ultra-high-energy photons and neutrinos in addition to cosmic rays.
This article provides a brief introduction to the physics of the radio emission of air showers, an overview of air-shower observatories using radio antennas, and highlights some of their recent results.
}


\maketitle

\section{Introduction}
This article reviews the status and recent progress regarding the radio detection of extensive air showers -- atmospheric particle cascades initiated by high energy cosmic particles.
It complements my other reviews on this topic~\cite{Schroder:2016hrv,Schroeder:2022icu,Schroder:2022vpg} and provides an updated overview of current and planned experiments featuring radio antennas for air-shower detection. 
For a more complete view of the topic including theory and the physics behind simulations of the radio signal, I also recommend reading reviews by other authors, such as \cite{Huege:2016veh,Alvarez-Muniz:2022uey}.
Particular emphasis in this review is given on recent experimental progress.

\section{Radio Emission of Air Showers}

The radio signal of air showers is mostly generated by the electrons and positrons of the electromagnetic shower component and beamed in the forward direction. 
The emission is coherent, i.e., the amplitude of the radio signal scales approximately linearly and the power quadratically with the number of electromagnetic particles, which itself is approximately proportional to the energy of the primary particle \cite{PierreAuger:2016vya}. 
The coherence conditions for the emission are optimal under the Cherenkov angle, which is of order $1^\circ$ in air, where the radio emission extends to the highest frequencies of a few GHz \cite{Smida:2014sia}.
As the shower front has a thickness of only a few meters, the radio emission is a short pulse, few ns to few $100$s of ns, and has a broad frequency spectrum with typical experiments operating in a band somewhere between 30 MHz and a few GHz.
Because the unavoidable Galactic radio noise decreases with frequency, the frequency band has impact on the signal-to-noise ratio \cite{BalagopalV:2017aan,Schroder:2023sam}.
Hence, the frequency band is an important aspect of the design of an antenna array and ideally is optimized for the science goals of an experiment.

\begin{figure}[t]
\centering
\includegraphics[width=0.69\linewidth]{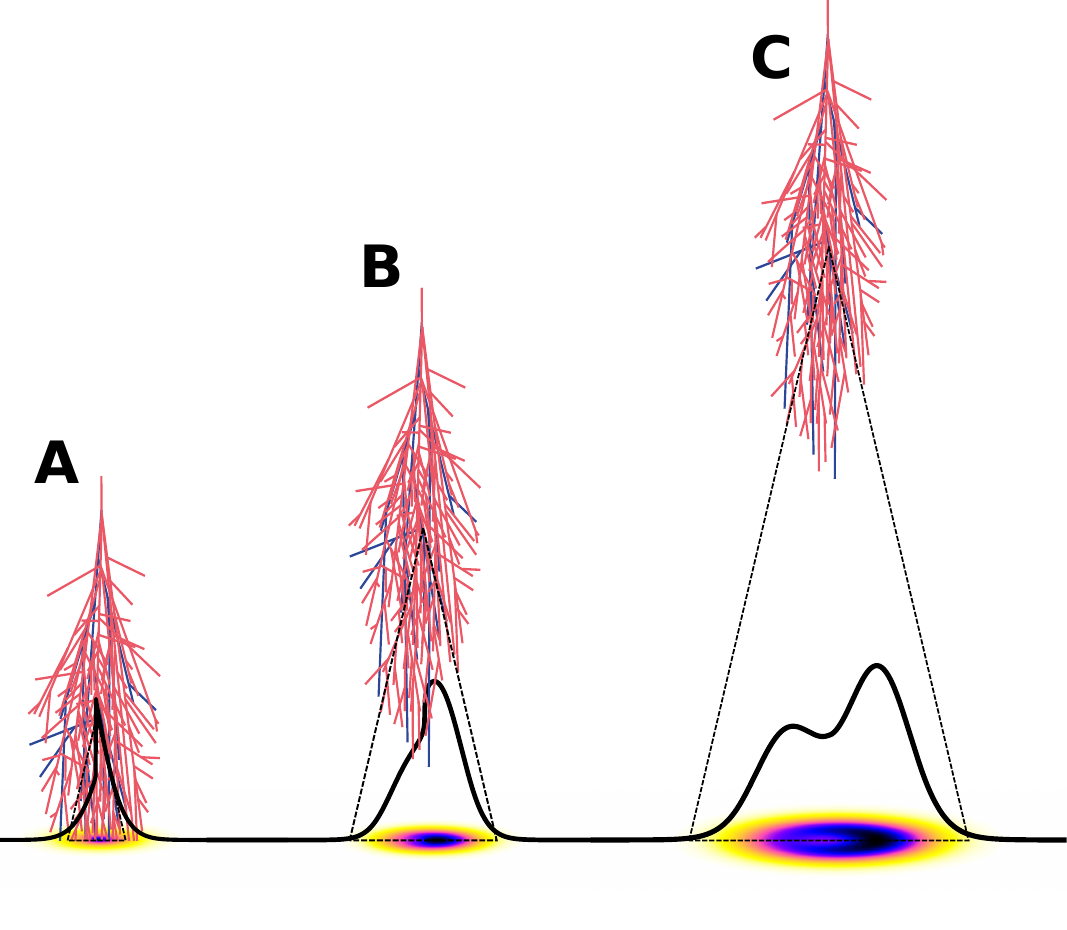}
\caption{Air showers with different depth of shower maximum (A to C, where A is closest to ground) and the lateral distribution of the energy fluence of their radio emission illustrated in two ways: through the darkness of the color of the-two dimensional footprint and through the one-dimensional lateral distribution. The asymmetry is caused by interference of geomagnetic and Askaryan emission. The dip in the center is visible only for distant shower maxima and results from the emission being enhanced at the Cherenkov angle. Reprinted from \cite{Glaser:2018byo} with permission from Elsevier.}
\label{fig_footprint}
\end{figure}

\begin{figure*}[t]
\centering
\includegraphics[width=0.99\linewidth]{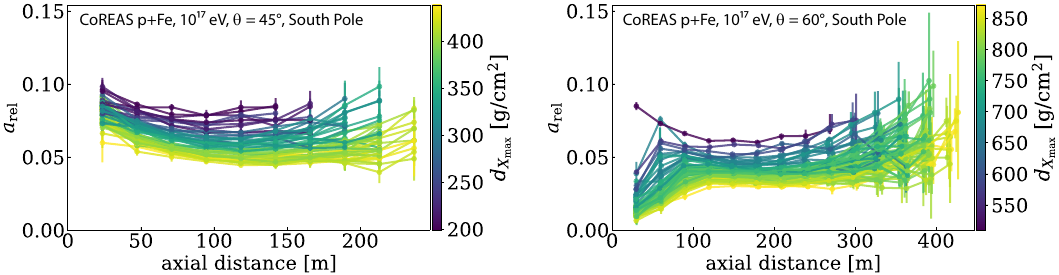}
\caption{Amplitude of the Askaryan emission $A$ relative to the geomagnetic $G$ emission for the ground level at the geographic South Pole for two different zenith angles as function of the distance from the shower axis and the distance to the shower maximum (determined with CoREAS simulations of the radio emission of air showers). The geomagnetic amplitude is corrected for the size of the geomagnetic Lorentz force with the plotted Askaryan fraction defined as $a_\mathrm{rel} = \frac{A}{G} \sin \alpha$, with $\alpha$ the angle between the Earth's magnetic field and the shower axis (figure modified from Ref.~\cite{Paudel:2022tbe}).}
\label{fig_AskaryanFraction}
\end{figure*}

The main mechanism of air-shower radio emission is the deflection of the electrons and positrons in the shower front by the Earth's magnetic field. 
This induces a time varying transverse current leading to radio emission linearly polarized in the $v \times B$ direction in the shower plane, where $v$ is the shower axis and $B$ the direction of the magnetic field. 
Therefore, the strength of the radio emission does not only depend on the zenith angle, but also on the azimuth angle of the air showers, which needs to be taken into account when modeling the aperture of radio arrays \cite{Lenok:2022gol}.
For very inclined showers and at high frequencies, the polarization pattern of geomagnetic emission becomes more synchrotron-like, which indicates that the curved tracks of the deflected electrons and positrons are important and not just the induction of the transverse current \cite{James:2022mea,Chiche:2024yos}.
Moreover, Askaryan emission due to the net charge excess in the shower front plays a small, but non-negligible role in air showers (see figure~\ref{fig_AskaryanFraction}).
Although it is an order of magnitude weaker than geomagnetic emission for most shower geometries, it interferes with geomagnetic emission, leading to an asymmetric lateral distribution of the radio signal on ground.
Figure \ref{fig_footprint} illustrates that asymmetric radio footprint on ground and its dependence on the distance to the shower maximum. 
The size of the radio footprint can be as small as $100-200\,$m in diameter for vertical showers or extend over several $10\,$km for near-horizontal air showers \cite{PierreAuger:2018pmw}.

When aiming at an accurate reconstruction of the air-shower direction, energy, and depth of maximum ($X_\mathrm{max}$), several subtle effects need to be taken into account to achieve the best possible accuracy:
For the direction, a plane-wave fit of the arrival times may achieve a resolution of $O({1^\circ})$, which is usually sufficient for cosmic-ray physics. 
Nonetheless, sub-degree resolution has been achieved when taking into account that the radio wavefront of air showers is approximately a hyperboloid \cite{Apel:2014usa,Corstanje:2014waa}, which in some cases can be further approximated by a spherical or conical wavefront.
For an accurate reconstruction of the energy of the air-shower, the azimuthal asymmetry of the radio footprint on ground is important \cite{Kostunin:2015taa,Tunka-Rex:2015zsa,PierreAuger:2016vya}, and one needs to be aware that the radiation energy is a not a direct measure for the total shower energy, but proportional to the size of the electromagnetic shower component. 
Depending on the method used, such as interferometry \cite{LOPES:2021ipp,Schoorlemmer:2020low} or template fitting \cite{Buitink:2014eqa,Bezyazeekov:2018yjw,IceCube:2023zne}, the shape of the radio wavefront and the lateral distribution are also important for an accurate reconstruction of $X_\mathrm{max}$.

\begin{figure*}[t]
\centering
\includegraphics[width=0.99\linewidth]{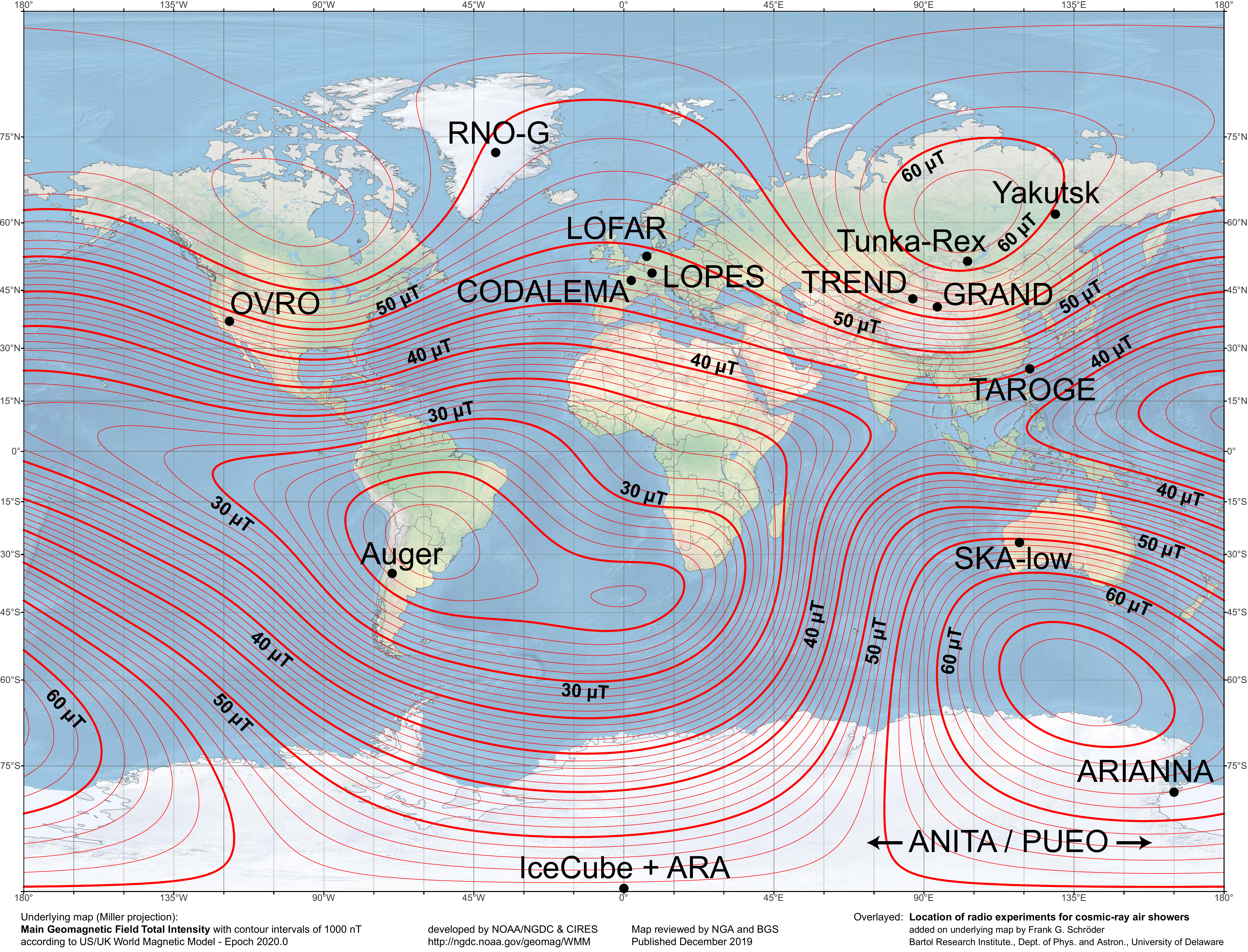}
\caption{Location of radio experiments for air-shower detection and the local strength of the geomagnetic field [see reference in the figure and in the text].}
\label{fig_magneticMap}
\end{figure*}

\section{Radio Experiments for Air Showers}
Radio pulses from air showers were first measured in the 1960's with analog experiments \cite{AllanReview}. 
While these measurements were sufficient to confirm the main features of geomagnetic radio emission, the accuracy obtained for important air-shower observables, such as the energy and depth of shower maximum could not compete with other techniques. 
This changed only with digital radio arrays combined with computational pipelines for data analysis \cite{PierreAuger:2011btp,Glaser:2019rxw,IceCube:2022dcd}, including modern methods, such as template fitting \cite{Buitink:2014eqa, Bezyazeekov:2018yjw, PierreAuger:2023rgk, PierreAuger:2023lkx}, and artificial neural networks \cite{Erdmann:2019nie,Bezyazeekov:2021sha,Rehman:2023jme}. 
Moreover, improved techniques for time \cite{Schroder:2010sa,PierreAuger:2015aqe} and amplitude calibration \cite{LOPES:2015eya,PierreAuger:2017xgp} of radio arrays have enabled a high data quality of current experiments.
Finally, state-of-the-art Monte Carlo simulation codes, such as CoREAS \cite{Huege:2013vt} and ZHAireS \cite{Alvarez-Muniz:2011ref}, describe the radio emission in sufficient detail for an accurate interpretation of measured data.

LOPES \cite{LOPES:2021ipp} and CODALEMA \cite{Ardouin:2005qe} were a first generation of digital radio arrays, providing several proof-of-principles, such as for digital radio interferometry of air showers \cite{LOPES:2005ipv} and for the sensitivity of radio measurements to the longitudinal shower development \cite{LOPES:2012xou}.\nocite{Petrov:2024yfb}
AERA \cite{PierreAuger:2023lkx}, LOFAR \cite{Schellart:2013bba}, TREND \cite{Ardouin:2010gz}, and Tunka-Rex \cite{Tunka-Rex:2015zsa} belong to a second generation that demonstrated a precision for the reconstructed shower energy and $X_\mathrm{max}$ comparable to the established air-fluorescence and air-Cherenkov techniques \cite{Buitink:2014eqa,PierreAuger:2016vya,Bezyazeekov:2018yjw,PierreAuger:2023lkx}. 
As enhancements of particle-detector arrays, radio detectors can thus enhance the measurement accuracy for the cosmic-ray energy and mass composition. 

Nowadays, a third generation of digital radio experiments has started aiming at cosmic-ray science in the energy range above $10^{16}\,$eV, with some experiments like the AugerPrime radio upgrade of the Pierre Auger Observatory under construction \cite{PierreAuger:2023gql}, and others planned, such as the IceCube-Gen2 surface array \cite{IceCubeGen2TDR}, with prototype stations in operation \cite{IceCube:2023zne}.
Following the example of LOFAR, also other antenna arrays built primarily for radio astronomy can also detect air showers in parallel to astronomical observations, in particular, OVRO \cite{Monroe:2019zkp} and the future SKA-low array \cite{Buitink:2023rso}.
Moreover, several experiments designed primarily to search for neutrino induced radio signals, also detected air showers, such as the ground-based radio arrays ARIANNA \cite{Barwick:2016mxm}, RNO-G \cite{RNO-G:2020rmc}, TAROGE \cite{TAROGE:2022soh}, BEACON \cite{Zeolla:2023phg}, and GRAND (with its GRANDproto300 protoype array for cosmic-ray air showers) \cite{GRAND:2018iaj}, and the balloon-borne radio probes ANITA \cite{ANITA:2010ect}, PUEO \cite{PUEO:2020bnn}, and POEMMA Balloon Radio (PBR) \cite{BATTISTI2024169819}.

\begin{table*}[t]
\centering
\caption{Selected digital antenna arrays used for air-shower detection, and the values of the international geomagnetic reference model (IGRF) for the experimental site: 
the zenith angle of the geomagnetic field, $\theta_\mathrm{geo} = 90^\circ - |\mathrm{inclination}|$, and the magnetic field strength $B_\mathrm{geo}$. 
The numbers of antennas, the approximate areas covered with antennas, and the nominal frequency bands are given. 
However, many experiments were used also in configurations different from the one stated in the table or feature additional antennas for other purposes than air-shower detection, and many analyses are based on subsets of antennas and smaller sub-bands.} \label{tab_experimentsSites}
\vspace{0.3cm}
\small
\begin{tabular}{lccrrccccr}
\hline
Experiment& Latitude & Longitude &Altitude& $\theta_\mathrm{geo}$ &$B_\mathrm{geo}$ &Number of& Area &Band\\
or location& &  & in m & in $^\circ$  &in \textmu T&antennas& in km\textsuperscript{2} & in MHz \\
\hline
LOPES & $49^\circ06$' N & $8^\circ26$' E & $110$ & $25.2^\circ$ & $48.1$ & $30$ & $0.04$ & $40-80$\\  
Yakutsk & $61^\circ 42$' N & $129^\circ 24$' E & $100$ & $13.5^\circ$ & $59.9$ & $6$ & $0.1$ & $32$\\  
CODALEMA & $47^\circ 23$' N & $2^\circ 12$' E & $130$ & $27.0^\circ$ & $47.5$ & $60$ & $1$ & $2-200$\\  
TREND & $42^\circ 56$' N & $86^\circ 41$' E & $2650$ & $26.2^\circ$ & $56.1$ & $50$ & $1.2$ & $50-100$\\ 
Auger & $35^\circ06$' S & $69^\circ30$' W & $1550$ & $53.0^\circ$ & $23.5$ & $153$ & $17$ & $30-80$\\ 
LOFAR & $52^\circ55$' N & $6^\circ52$' E & $5$ & $22.0^\circ$ & $49.7$ & $\approx 300$ & $0.2$ & $10-240$\\ 
Tunka-Rex & $51^\circ49$' N & $103^\circ04$' E & $675$ & $18.0^\circ$ & $60.4$ & $63$ & $1$ & $30-80$\\  
South Pole & $90^\circ$ S & - & $2834$ & $18.0^\circ$ & $54.4$ & few & - & various\\ 
ARIANNA & $78^\circ45$' S & $165^\circ00$' E & $400$ & $10.0^\circ$ & $62.1$ & $32$ & $5$ & $50-1000$\\ 
SKA-low & $26^\circ 41$' S & $116^\circ 38$' E & $370$ & $29.9^\circ$ & $55.6$ & $60,000$ & $1$ & $50-350$\\ 
RNO-G & $72^\circ 35$' N & $38^\circ 28$' W & $3216$ & $18.9^\circ$ & $54.8$ & $105$ & $50$ & $80-650$\\ 
OVRO & $37^\circ 14$' N & $118^\circ 17$' W & $1222$ & $38.4^\circ$ & $48.0$ & $256$ & $0.04$ & $30-80$\\ 
TAROGE & $24^\circ 17$' N & $121^\circ 44$' E & $1000$ & $53.8^\circ$ & $45.2$ & $4$ & $-$ & $180-350$\\ 
BEACON & $37^\circ 35$' N & $118^\circ 14$' W & $1180$ & $38.1^\circ$ & $48.2$ & $4$ & $-$ & $30-80$\\ 
GRAND\tiny{proto300} & $41^\circ$ N & $94^\circ$ E & $1300$ & $37.9^\circ$ & $56.7$ & $230$ & $200$ & $50-200$\\ 
\hline
\end{tabular}
\end{table*}

Figure~\ref{fig_magneticMap} and Table~\ref{tab_experimentsSites} provide an overview of the location and characteristics of several of these radio experiments for air-shower detection. 
A table with the abbreviated and full names of the experiments can be found at the end of this article.

Many of these radio arrays are triggered by other air-shower detectors, such as arrays of particle detectors, and share a common infrastructure for data taking. 
Self-triggering, which is required for stand-alone radio detection, has been demonstrated by several experiments \cite{ANITA:2010ect,Ardouin:2010gz, Monroe:2019zkp,PierreAuger:2012gwg}. 
However, self-triggering increases the detection threshold compared to external air-shower triggers and is mostly important for radio detectors aiming at huge exposure for ultra-high-energy neutrinos searches.
For many cosmic-ray science goals, not only statistics, but also high accuracy is important, which is more easily achieved by hybrid detection of air showers with several techniques.
Thus, an external trigger comes almost for free in air-shower arrays featuring coincident detection of air showers with particle detectors and radio antennas.

\begin{figure*}[t]
\centering
\includegraphics[width=0.99\linewidth]{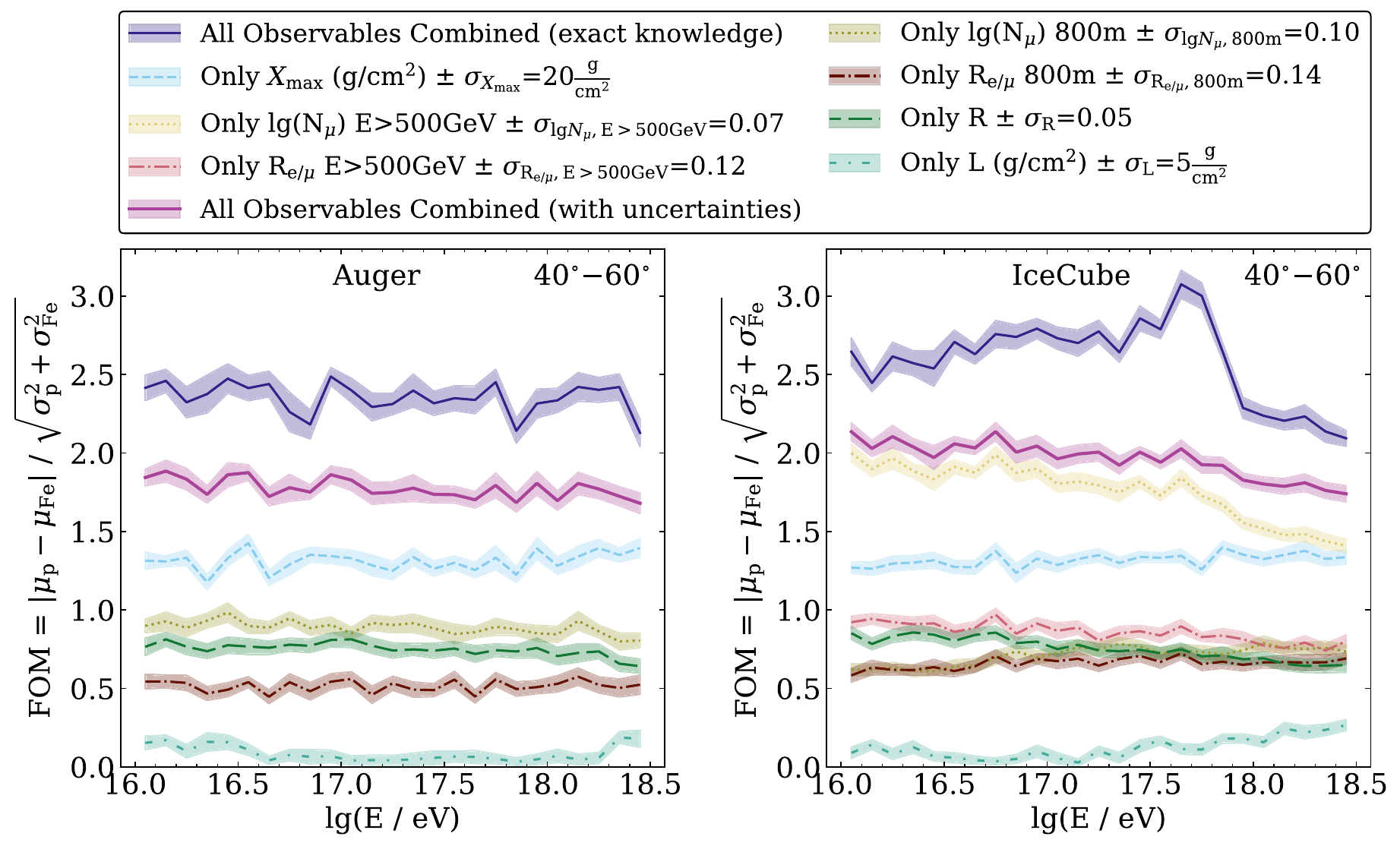}

\caption{Figure of merit (FOM) for the event-by-event sensitivity of various air-shower observables and their combination to separate air showers initiated by protons and iron nuclei. The results are for CORSIKA simulations with Sibyll 2.3c and 2.3d in the zenith angle range of $40^\circ - 60^\circ$ for the sites of the Pierre Auger Observatory and IceCube, respectively. $X_\mathrm{max}$ and the muon number $N_\text{\textmu}$ provide the highest separation power, where muons for the Auger site are muons at $800\,$m axis distance at the surface and for IceCube high-energy muons as detectable by the deep detector. $R_{\text{e}/\text{\textmu}}$ is the electron-muon ratio at ground. $X_\mathrm{max}$, $R$, and $L$ are parameters of the Gaisser-Hillas function describing the longitudinal shower profile. No detector simulation has been done, but some measurement uncertainties have been included as stated in the legend (figure modified from Ref.~\cite{Flaggs:2023exc}).}

\label{fig_massSensitivity}
\end{figure*}

\section{Future Developments}
Several future projects plan to use radio instrumentation, mostly to increase the accuracy of air-shower measurements. 
In particular, increasing the accuracy for the rigidity of cosmic-ray nuclei is important to search for the sources of the most energetic Galactic and extragalactic cosmic rays through mass-sensitive anisotropy \cite{Schroder:2019rxp,Coleman:2022abf}.
In addition to an accurate energy determination, which can be provided by radio detectors, this primarily requires event-by-event mass sensitivity. Provided that the shower energy is known, the two most mass sensitive shower observables are the the depth of shower maximum, $X_\mathrm{max}$, and the muon number $N_\text{\textmu}$.
Therefore, air-shower arrays featuring both radio and muon detection are ideal for this purpose \cite{Holt:2019fnj,Flaggs:2023exc}.

In addition to further efforts to increase the accuracy of radio measurements through better calibration and analysis techniques, a number of new projects are planned for the next decade relying on radio detection, some of them making use of coincident shower measurements of radio and muon detectors:

\begin{figure*}[t]
\centering
\includegraphics[angle=90,width=0.48\linewidth]{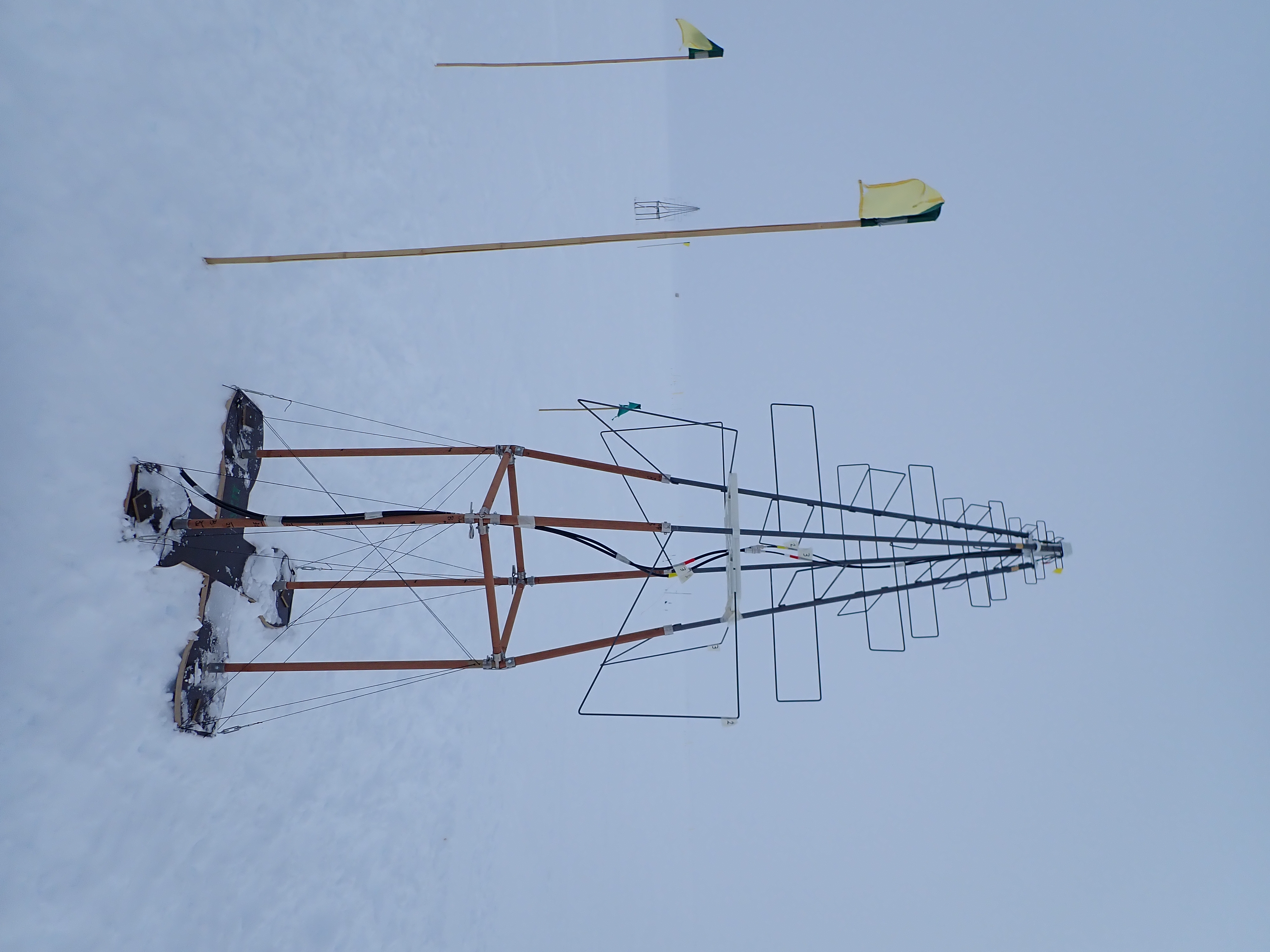}
\hfill
\includegraphics[angle=90,width=0.48\linewidth]{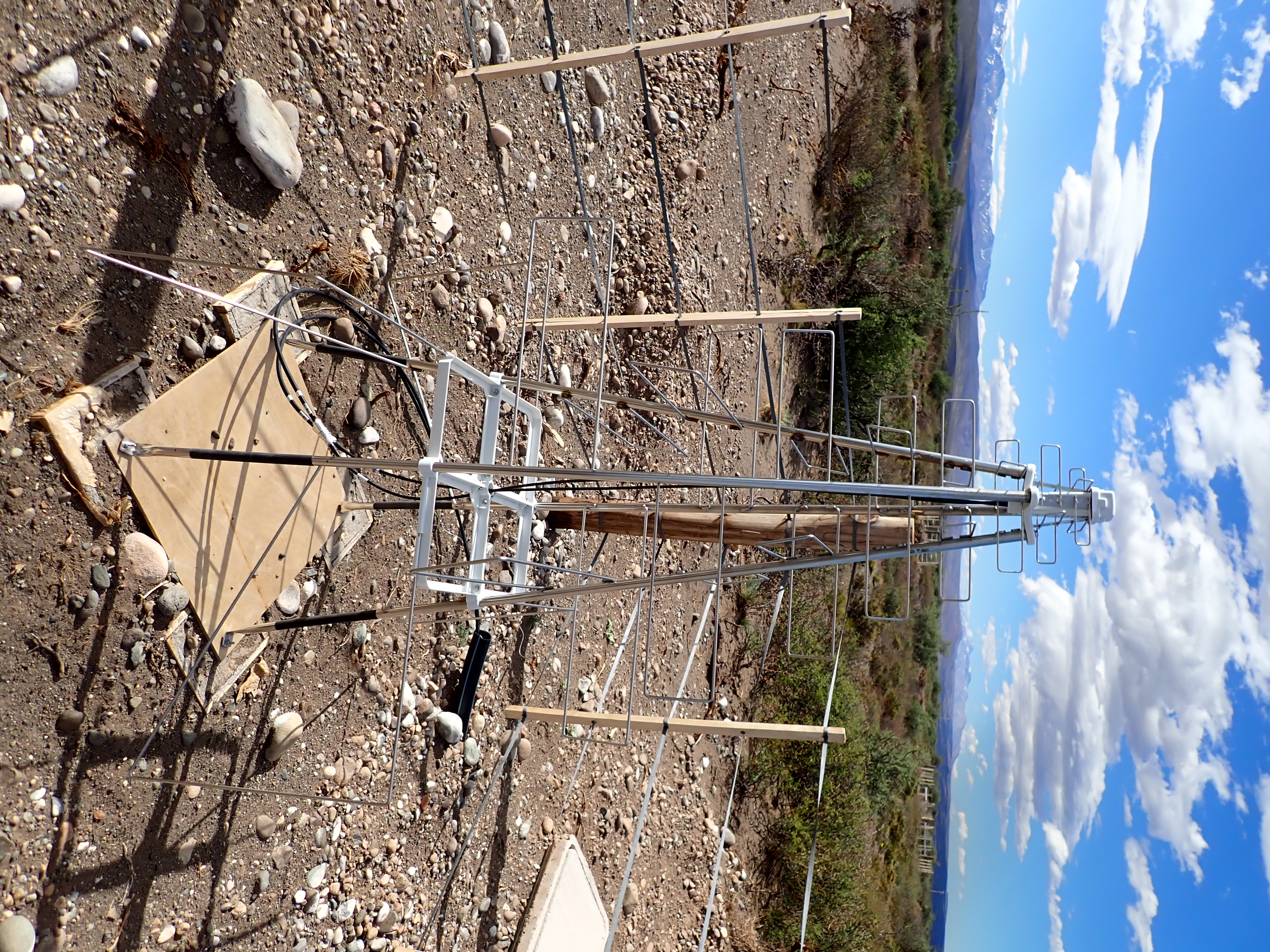}
\caption{Photos of SKALA v2 antennas used in prototype stations of the IceCube-Gen2 surface array at IceCube at the South Pole (left) and at the Pierre Auger Observatory in Argentina (right).}
\label{fig_SKALAs}
\end{figure*}

IceCube-Gen2 \cite{IceCubeGen2TDR} will continue a long tradition of cosmic-ray physics at the South Pole \cite{Soldin:2023lbr}, as cosmic rays in the energy range of the Galactic-to-extragalactic transition will complement the primary science case of neutrino astronomy. 
The surface array will feature scintillation panels and radio antennas for air showers \cite{Schroder:2023lux}, which in combination with high-energy muons measured by an optical array deep in the ice will provide high per-event mass sensitivity.

GCOS will consist of air-shower arrays at several sites, together increasing the exposure of current ultra-high-energy cosmic-ray observatories by an other of magnitude and at the same time featuring per-event mass sensitivity \cite{AlvesBatista:2023lqg}. 
At least one site will likely feature radio antennas, possibly for calibration or as enhancement of water-Cherenkov detectors following the concept of AugerPrime, the upgrade of the Pierre Auger Observatory.
Among several other improvements of the Pierre Auger Observatory, such as new electronics and scintillation panels, a radio antenna has been added to each surface detector to provide a radio measurement of highly inclined air showers \cite{PierreAuger:2023gql}.

There are also science cases in cosmic-ray physics for future radio arrays without muon detectors. 
Examples are the huge exposure that can be provided by GRAND \cite{GRAND:2018iaj}. 
As GCOS, GRAND will feature several sites, with a total area even larger than GCOS. 
Although GRAND will be optimized for neutrino detection, it will also collect a huge statistics of ultra-high-energy cosmic rays, and common sites between GRAND and GCOS can leverage synergies.

For the PeV-to-EeV energy range, SKA-low will provide the option to observe air showers in parallel to radio astronomy. 
The huge number of several $10,000$s of antennas will enable measuring the radio emission of air showers in unprecedented detail, including the shape parameters $L$ and $R$ for the width and asymmetry of the longitudinal shower profile, i.e., the number of particles as a function of atmospheric depth $X$:
\begin{equation*}
N(X) = N_\mathrm{max} \left(1 + \frac{R (X - X_\mathrm{max})}{L} \right)^{R^{-2}} \exp{\left( - \frac{X-X_\mathrm{max}}{L R} \right)}    
\end{equation*} 
While $L$ and $R$ do not feature significant event-by-event mass separation power (cf.~figure~\ref{fig_massSensitivity}), they provide a novel approach to test hadronic interaction models and to statistically determine the proton and Helium fractions among the primary cosmic-ray particles \cite{Buitink:2021pkz}.
Last but not least, there are significant synergies among these experiments, e.g., the SKALA antennas developed for SKA-low \cite{SKALAV2} are also the reference design for the IceCube-Gen2 surface array, and prototype stations with SKALA v2 antennas have been deployed not only at the South Pole, but also at the Telescope Array and at the Pierre Auger Observatory (see figure~\ref{fig_SKALAs}).

\section{Conclusion}
Digital antenna arrays have been shown to feature an accuracy for the arrival direction, energy, and $X_\mathrm{max}$ competitive with established detection techniques for air showers. 
This achievement has been enabled by progress in the instrumentation and its calibration, an improved understanding of the radio emission of air showers, and by sophisticated computational analysis techniques.
While progress on all of these aspects continues, the level achieved is already sufficient to make radio detection an essential technique for cosmic-ray physics of the coming decade.
We are therefore seeing a number of experiments targeting both particle physics and astrophysics questions related to the most energetic Galactic and extragalactic cosmic rays with the help of radio antennas.
\\
~
\\
\clearpage
\noindent
\textbf{Abbreviated and full names of mentioned experiments / observatories:}
\\
\begin{tabular}{ll}
AERA & Auger Engineering Radio Array\\
ANITA & Antarctic Impulsive Transient Antenna\\
ARA & Askaryan Radio Array\\
ARIANNA & Antarctic Ross Ice-Shelf Antenna Neutrino Array\\
Auger & Pierre Auger Observatory\\
BEACON & Beamforming Elevated Array for COsmic Neutrinos\\
CODALEMA & COsmic ray Detection Array with Logarithmic\\
& ElectroMagnetic Antennas\\
GCOS & Global Cosmic-Ray Observatory\\
GRAND & Giant Radio Array for Neutrino Detection\\
IceCube-Gen2 & IceCube Neutrino Observatory -- Generation 2\\
LOFAR & Low-Frequency Array\\
LOPES & LOFAR Prototype Station\\
OVRO & Owens Valley Radio Observatory\\
PBR & POEMMA-Balloon with Radio\\
POEMMA & Probe of Extreme Multi-Messenger Astrophysics\\
PUEO & Payload for Ultrahigh Energy Observations\\
RNO-G & Radio Neutrino Observatory in Greenland\\
SKA & Square Kilometer Array\\
TAROGE & Taiwan Astroparticle Radiowave Observatory\\
 & for Geo-synchrotron Emissions\\
TREND & Tianshan Radio Experiment for Neutrino Detection\\
Tunka-Rex & Tunka Radio Extension\\
\end{tabular}
\\
\\
Caveat: In some cases the full name does not reflect later stages of the projects, e.g., because of additional locations or antenna types used.

\backmatter

\bmhead{Acknowledgements}
The author thanks the organizers of the Radio Neutrino and Cosmic Rays Astronomy Workshop for the opportunity to deliver a remote lecture on 19 April 2024. The research of the author is supported by several funding agencies:
NASA EPSCoR award \#80NSSC22M0222, U.S. National Science Foundation awards \#2019597, \#2046386, and \#2209483, Sloan Research Foundation.
This project has received funding from the European Research Council (ERC) under the European Union’s Horizon 2020 research and innovation programme (grant agreement No 802729).
This research was supported in part through the use of Information Technologies (IT) resources at the University of Delaware, specifically the high-performance computing resources.


\bibliography{bibliographyReview}

\end{document}